\documentclass[letterpaper,twocolumn,10pt]{article}
\usepackage{usenix_style}

\makeatletter
\def\@maketitle{\newpage
  \vbox to 1.0in{%
    \vspace*{\fill}%
    \begin{center}%
      {\Large\bfseries \@title \par}%
      \vskip 0.1in minus 0.05in
      {\large\itshape
      \lineskip .3em
      \begin{tabular}[t]{c}\@author\end{tabular}\par}%
    \end{center}%
    \vspace*{\fill}%
  }%
}
\makeatother

\usepackage{tikz}
\usepackage{amsmath}

\usepackage{filecontents}
\usepackage{soul}
\usepackage{algorithm}  
\usepackage{algorithmicx}  
\usepackage[noend]{algpseudocode}
\usepackage{eqparbox}
\usepackage{xcolor}

\usepackage{graphicx}
\usepackage{subfigure}
\usepackage{enumitem}

\usepackage[hang,flushmargin]{footmisc}

\usepackage[skip=0pt]{caption} 
\usepackage{tabularx}
\usepackage{booktabs}
\usepackage{hhline}

\usepackage{soul}

\usepackage{url}
\usepackage{multirow}
\usepackage{comment}
\usepackage[explicit]{titlesec}

\usepackage{pifont}
\usepackage{adjustbox}
\usepackage{todonotes}
\usepackage{latexsym}

\usepackage{float}

\usepackage[normalem]{ulem}
\begin{document}

\title{FlexKV: Flexible Index Offloading for Memory-Disaggregated Key-Value Store}
\author{
Zhisheng Hu$^{1}$ \quad
Jiacheng Shen$^{2}$ \quad
Ming-Chang Yang$^{1}$ \\
$^{1}$The Chinese University of Hong Kong \quad
$^{2}$Duke Kunshan University
}

\maketitle

\begin{abstract}
Disaggregated memory (DM) is a promising data center architecture that decouples CPU and memory into independent resource pools to improve resource utilization. 
Building on DM, memory-disaggregated key-value (KV) stores are adopted to efficiently manage remote data. 
Unfortunately, existing approaches suffer from poor performance due to two critical issues: 1) the overdependence on one-sided atomic operations in index processing, and 2) the constrained efficiency in compute-side caches. 
To address these issues, we propose \textbf{\textsf{FlexKV}}, a memory-disaggregated KV store with index proxying. 
Our key idea is to dynamically offload the index to compute nodes, leveraging their powerful CPUs to accelerate index processing and maintain high-performance compute-side caches.
Three challenges have to be addressed to enable efficient index proxying on DM, \textit{i.e.}, the load imbalance across compute nodes, the limited memory of compute nodes, and the expensive cache coherence overhead.
\textsf{FlexKV} proposes: 
1) a \textit{rank-aware hotness detection} algorithm to continuously balance index load across compute nodes, 2) a \textit{two-level CN memory optimization} scheme to efficiently utilize compute node memory, and 3) an \textit{RPC-aggregated cache management} mechanism to reduce cache coherence overhead.
The experimental results show that \textsf{FlexKV} improves throughput by up to $2.94\times$ and reduces latency by up to $85.2\%$, compared with the state-of-the-art memory-disaggregated KV stores.
\end{abstract}

\section{Introduction}\label{section_1_intro}
Disaggregated memory (DM), a promising data center architecture proposed to improve resource efficiency, has gained significant attention from both industry~\cite{gaussdb, polarcxl, polardb, aurora, polardb-mp} and academia~\cite{infiniswap, legoos, dmorigin, sysimply}. 
DM breaks the boundaries of monolithic servers by decoupling CPUs and memory into the compute pool (compute nodes, CNs) and the memory pool (memory nodes, MNs), connected via high-performance networks like RDMA~\cite{rdmaspec, designguideatc} and CXL~\cite{cxlspec, cxlshm, cxlpond, oasis}. 
This architecture enables CPUs and memory to be flexibly allocated, enhancing overall resource efficiency.

Key-value (KV) stores are emerging applications on DM~\cite{fusee,dinomo,aceso,clover,swarm}.
Scalability, \textit{i.e.}, the ability to serve more clients without significant performance degradation, is a key requirement for all memory-disaggregated KV stores. 
However, existing approaches still struggle to achieve this goal due to the following two primary issues. 

\textbf{The overdependence on one-sided atomic operations in index processing.} 
Existing memory-disaggregated KV stores extensively rely on one-sided atomic operations (\textit{i.e.}, \texttt{CAS} and \texttt{FAA}) to resolve concurrency conflicts during index processing~\cite{designguideacm}.
Unfortunately, the extensive number of one-sided atomic operations generates a severe performance bottleneck on DMs implemented with both RDMA and CXL.
Specifically, for RDMA-based DM, atomic operations with more complex semantics consume more processing power than read or write operations~\cite{rdmaunderstand}, making the processing units of RDMA network interface cards (RNICs) a bottleneck.
Meanwhile, for CXL-based DM, atomic operations hinder system performance due to the expensive cross-node cache coherence protocol (\textit{i.e.}, \texttt{CXL.cache} in CXL 3.0)~\cite{cxlspec,cxlhydrarpc,cxldemysify,cxl-character}.

\textbf{The constrained efficiency in compute-side caches.} 
Due to the significant performance gap between remote and local memory, using the local memory of CNs as caches is a common practice in many memory-disaggregated KV stores~\cite{dinomo, dex, smartluo}.
To reduce cross-node cache coherence overhead, existing designs either cache only the index (\textit{i.e.}, addresses of KV pairs)~\cite{fusee,clover,chimeluo} or cache both the index and KV pairs but partition the ownership of cached data to individual CNs~\cite{dex,dinomo}.
However, both approaches are suboptimal in terms of system performance.
The former requires at least one remote memory access per request even on cache hits, while the latter introduces additional overhead from forwarding KV requests to designated nodes that own the requested data.

In this paper, we propose to offload part of the index to the compute pool to address the above issues simultaneously.
Specifically, by offloading the index, CNs can help with index processing, converting expensive one-sided atomic operations into local CPU operations.
Furthermore, during the offloaded index processing, CNs can efficiently cache KV pairs, enhancing compute-side cache efficiency.
However, three challenges have to be addressed to achieve this idea in practice.

\textbf{(1) The load imbalance across compute nodes.}
Index partitions may experience vastly different access frequencies due to workload skewness. When offloading them to CNs with static assignment, hot partitions can overload certain CNs while leaving others underutilized, degrading overall performance. Moreover, workload patterns are dynamic in production environments~\cite{pegasus, netcache, facebookworkload, twitter-workload}, requiring continuous hotness tracking and dynamic reassignment. However, frequent reassignments cause significant overhead, making it challenging to balance load effectively.

\textbf{(2) The limited memory of compute nodes.}
CNs have abundant CPUs but very limited memory~\cite{fusee, aifm, chimeluo}, making it impractical to store the entire index locally in CNs, especially for large-scale production datasets~\cite{rocksdb, twitter-workload, facebookworkload}. Furthermore, offloading the index, whether fully or partially, to CNs introduces additional memory overhead, particularly from metadata needed for index operations. This further strains the already limited CN memory, interfering with user request processing and degrading system performance.

\textbf{(3) The expensive coherence overhead of caching KV pairs.}
When KV pairs are cached across multiple CNs, maintaining cache coherence becomes crucial for ensuring data consistency and correctness in the memory-disaggregated KV store. 
However, implementing a coherence protocol across network-connected CNs introduces significant overhead. 
First is the routine metadata maintenance overhead, \textit{i.e.}, the system must continuously manage and update cache metadata, such as the cache directory. 
Second is the bursty write invalidation overhead, \textit{i.e.}, updating a KV pair that is cached on multiple CNs triggers invalidations across those nodes. 
These operations substantially amplify the number of remote accesses, making caching KV pairs expensive.

To address the above challenges, we propose \textbf{\textsf{FlexKV}}, a memory-disaggregated KV store with index proxying that can dynamically offload the index to the compute pool and maintain efficient compute-side caches for KV pairs.
First, to address load imbalance, \textsf{FlexKV} adopts a \textit{rank-aware hotness detection} algorithm that continuously tracks partition access frequencies and triggers reassignments only when workload imbalance becomes significant.
Second, to overcome limited CN memory, \textsf{FlexKV} employs a \textit{two-level CN memory optimization} scheme: it uses \textit{slot-resolved index RPCs} to eliminate heavy key-to-slot mappings on CNs, and introduces an \textit{adaptive index-cache splitting} technique that efficiently partitions the limited CN memory between index and cache to maximize performance.
Finally, to reduce cache coherence overhead, \textsf{FlexKV} leverages an \textit{RPC-aggregated cache management} mechanism that piggybacks cache metadata maintenance (\textit{e.g.}, directory updates and hotness tracking) onto index processing, amortizing metadata maintenance costs while enabling selective caching of read-intensive KV pairs.

We implement \textsf{FlexKV} and evaluate its performance with YCSB~\cite{ycsbworkload} and Twitter~\cite{twitter-workload} workloads. The results show that \textsf{FlexKV} improves throughput by up to $2.94\times$ and reduces latency by up to $85.2\%$, compared with Clover~\cite{clover}, FUSEE~\cite{fusee}, and Aceso~\cite{aceso}, three state-of-the-art memory-disaggregated KV stores. 

In summary, this paper makes the following contributions:
\begin{itemize}[leftmargin=20pt, itemsep=0.5pt, parsep=0pt, topsep=1pt]
  \item We identify two key issues in existing approaches: 1) the overdependence on one-sided atomic operations in index processing, and 2) the constrained efficiency in compute-side caches. We propose that offloading the index to the compute pool can address both issues.
  \item We present \textsf{FlexKV}, a memory-disaggregated KV store with index proxying. By dynamically offloading the index to the compute pool, \textsf{FlexKV} prevents index processing from becoming the performance bottleneck and achieves efficient compute-side caching for KV pairs.
  \item We implement \textsf{FlexKV} and conduct a thorough evaluation, demonstrating the effectiveness of our design.
\end{itemize}

\section{Background}
\subsection{Disaggregated Memory}

\begin{figure}[tp]
    \centering
    \setlength{\abovecaptionskip}{0.1em}
    \includegraphics[width=\linewidth]{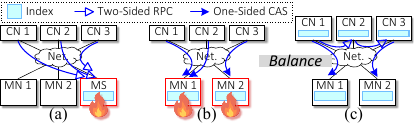}
    \captionsetup{font=small}
    \caption{Index deployment in memory-disaggregated KV stores. }
    \label{fg-bg-index}
\end{figure}

In traditional data centers, the unit of deployment is monolithic servers, where CPUs and memory are allocated in fixed proportions.
Such a server-centric approach results in significant resource underutilization~\cite{alibabatrace, borgtrace, wholimits}. 
On disaggregated memory (DM), monolithic servers are decoupled into compute nodes (CNs) and memory nodes (MNs). 
CNs have ample CPUs but limited memory, while MNs have abundant memory but only a few wimpy CPUs. 
Such a separation enables flexible, on-demand allocation of CPU and memory resources by adding the appropriate type of node to the cluster, thus achieving better resource utilization~\cite{concordia, zombie, legoos, mind, infiniswap}.

In a memory-disaggregated data center, CNs and MNs are connected through high-performance networks such as RDMA~\cite{rdmaspec} and CXL~\cite{cxlspec}.
Since CXL 3.0 devices are still under development, this paper focuses on RDMA-based DM, but the techniques discussed also consider future CXL-based DM (\S\ref{section_discuss_cxl}).
RDMA provides applications with one-sided and two-sided verbs for accessing remote data. One-sided operations (\textit{e.g.}, \texttt{WRITE}, \texttt{READ}) enable direct memory access on remote nodes without remote CPU involvement, with one-sided atomic operations like compare-and-swap (\texttt{CAS}) and fetch-and-add (\texttt{FAA}) for handling concurrent conflicts\cite{designguideacm}. Two-sided operations (\textit{e.g.}, \texttt{SEND}, \texttt{RECV}) are similar to traditional sockets, requiring CPU participation at both ends.

\subsection{Memory-Disaggregated KV Stores}

\subsubsection{Index Deployment on DM.}\label{sec-bg-index-deploy}

Based on the location of the index, existing memory-disaggregated KV stores can be classified into two categories.

\textit{(1) In monolithic metadata servers.} 
For example, in KV stores like Clover~\cite{clover} and SWARM-KV~\cite{swarm}, while KV pairs are stored in MNs, the index is stored in additional monolithic metadata servers (MS), as shown in Figure~\ref{fg-bg-index}(a). 
This design leads to significant resource waste on the monolithic servers managing the index~\cite{fusee}. 
Moreover, as all cluster-wide index operations are concentrated on a few monolithic servers, they can easily become the performance bottleneck.

\textit{(2) In memory nodes.} 
Many memory-disaggregated KV stores choose to distribute the index across multiple MNs (\textit{e.g.}, Aceso~\cite{aceso}, FUSEE~\cite{fusee}, DINOMO~\cite{dinomo}), as shown in Figure~\ref{fg-bg-index}(b). 
Such an approach eliminates the need for monolithic servers and achieves full disaggregation, thus enhancing resource utilization. 
However, these approaches heavily rely on one-sided operations to read/write remote memory in MNs. 
The performance of this design is directly tied to the limited processing power of RNICs in the MNs, especially for high-cost one-sided atomic operations. 
Since the number of RNICs for each MN is fixed and cannot be adjusted dynamically, they become a severe performance bottleneck under high concurrency, undermining disaggregated memory's core promise: independent scaling of compute and memory.

\textit{\uline{\textbf{Our choice}}: In both memory nodes and compute nodes.} 
In this paper, we propose to distribute the index across both MNs and CNs (\textit{i.e.}, dynamically offloading a portion of the index from MNs to CNs), as shown in Figure~\ref{fg-bg-index}(c). 
This design allows the index processing performance to scale with the expansion of either MNs or CNs, breaking the bottleneck.

\begin{figure}[tp]
    \centering
    \setlength{\abovecaptionskip}{0.1em}
    \includegraphics[width=\linewidth]{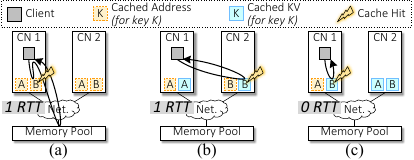}
    \captionsetup{font=small}
    \caption{Compute-side caching in memory-disaggregated KV stores. A client in CN 1 attempts to fetch the KV pair with key \textit{B}. }
    \label{fg-bg-cache}
\end{figure}

\subsubsection{Compute-Side Caching on DM.}\label{sec-bg-cache-deploy}

Based on the design of compute-side caches, existing approaches can also be classified into two categories.

\textit{(1) Address-only cache.} To minimize the overhead of maintaining cache coherence, many memory-disaggregated KV stores cache (\textit{e.g.}, Aceso~\cite{aceso}, FUSEE~\cite{fusee}, Clover~\cite{clover}) only the index (\textit{i.e.}, addresses of KV pairs), as shown in Figure~\ref{fg-bg-cache}(a). 
When reading a KV pair through a cached address, the client verifies its correctness by checking a valid bit in the KV pair header. 
If the KV pair is invalid, the client invalidates the cached address. 
However, even on cache hits, this address-only caching strategy requires at least one remote memory access to fetch the actual KV pair, limiting performance.

\textit{(2) Hybrid cache with ownership partitioning.} 
Some memory-disaggregated KV stores adopt a hybrid caching strategy that caches both the index and KV pairs (\textit{e.g.}, DINOMO~\cite{dinomo}, DEX~\cite{dex}), as shown in Figure~\ref{fg-bg-cache}(b).  
To minimize the overhead of ensuring cache coherence, they partition the ownership of data among CNs, where each CN handles the KV requests only for a given key range. 
Such an approach eliminates the need to invalidate caches on other CNs when one CN updates data. 
However, they adopt a client-CN-MN three-tier architecture. 
When a client initiates a KV request, if the CN responsible for the key is not local, the client must first forward the request to the designated CN, whose CPU then processes the subsequent operations and returns the result. 
As KV requests cannot be served by arbitrary CNs, this architecture exacerbates load imbalance and limits system performance due to the additional network overhead of forwarding, as demonstrated in our experiment (\S\ref{section_exp_op_cost}).

\textit{\uline{\textbf{Our choice}}: Hybrid cache with coherent sharing.} 
In this paper, we propose a hybrid caching strategy where CNs share cached 
data with coherence maintained via a directory-based protocol, as shown in Figure~\ref{fg-bg-cache}(c).
This design allows a KV pair to be cached by any CN, eliminating the MN-access overhead in address-only caching. Simultaneously, a KV request can be served by any CN, avoiding the load imbalance and forwarding overhead inherent to ownership partitioning.
Finally, we adopt a directory-based coherence protocol rather than a snoop-based protocol~\cite{ccnuma} because snoop-based protocols rely on broadcast messages, which limits system scalability and degrades performance as the number of CNs increases. 
\section{Motivation}

In the following sections, we provide a deeper analysis of the differences between different schemes of index operations (\textit{i.e.}, one-sided CAS vs. proxied CAS), as well as between remote data accesses and local cache hits (\textit{i.e.}, one-sided read vs. local read), to further motivate our design choices.

\subsection{One‑Sided CAS vs. Proxied CAS}

\begin{figure}[tbp]
    \centering
    \begin{minipage}[t]{0.49\linewidth}
      \centering
      \includegraphics[width=\textwidth]{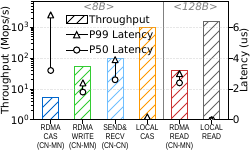}
      \captionsetup{font=small}
      \caption{The throughput and latency of different operations.}
      \label{fig-pre-cas-tpt}
    \end{minipage}
    \hspace{0em}
    \begin{minipage}[t]{0.49\linewidth}
      \centering
      \includegraphics[width=\textwidth]{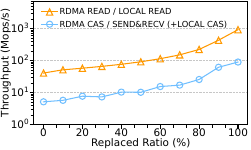}
      \captionsetup{font=small}
      \caption{The effect of varying the ratio of replaced operations.}
      \label{fig-pre-read-tpt}
    \end{minipage}
\end{figure}

Existing memory-disaggregated KV stores typically rely on one-sided atomic operations to resolve concurrency conflicts. 
As mentioned in \S\ref{section_1_intro}, these one-sided atomic operations are more complex and incur greater overhead on the processing units within RNICs.
In these KV stores, frequent invocation of one-sided atomic operations can easily make the MN RNICs a performance bottleneck.
In contrast, letting the CN CPUs handle concurrency conflicts using two-sided RPCs (\texttt{RDMA\_SEND\&RECV}) together with local atomic instructions (\texttt{LOCAL\_CAS}) can significantly improve performance, as they are more lightweight and avoid accessing the MNs. 

Figure~\ref{fig-pre-cas-tpt} shows the throughput of \texttt{RDMA\_CAS}, \texttt{RDMA\_WRITE}, \texttt{RDMA\_SEND\&RECV}, and \texttt{LOCAL\_CAS}, all at 8B granularity, with 200 clients (\textit{i.e.}, the same setup as in §\ref{sec-exp-setup}), as well as their P50 and P99 latencies in the single-client-per-CN case.
Among these operations, \texttt{RDMA\_CAS} and \texttt{RDMA\_WRITE} are issued from CNs to MNs, while \texttt{RDMA\_SEND\&RECV} occurs between CNs.
\texttt{RDMA\_WRITE}, \texttt{RDMA\_SEND\&RECV}, and \texttt{LOCAL\_CAS}  achieve $10.1\times$, $19.5\times$, and $177.1\times$ higher throughput than \texttt{RDMA\_CAS}, respectively, while also exhibiting lower latency, particularly \texttt{LOCAL\_CAS}. Notably, this performance gap holds across RNIC generations (\textit{e.g.}, ConnectX-4~\cite{hybrid-better}/5~\cite{rdmaturing}/6~\cite{rdmaunderstand}), since the relatively high overhead of one-sided atomics remains unchanged.
These results highlight the superior efficiency of these operations compared to the inefficient \texttt{RDMA\_CAS}, suggesting that replacing \texttt{RDMA\_CAS} with them under high load can provide substantial performance improvement. 
Figure~\ref{fig-pre-read-tpt} shows the throughput clearly increases when gradually replacing \texttt{RDMA\_CAS} with the combination of \texttt{RDMA\_SEND\&RECV} and \texttt{LOCAL\_CAS}, validating this observation.

Based on the above observation and to address the issue described in \S\ref{sec-bg-index-deploy}, \textsf{FlexKV} proposes the index proxying that dynamically offloads the index to CNs. 
Each CN can act as the proxy for an exclusive set of index partitions, managing these partitions in its local memory via \texttt{LOCAL\_CAS}. All index operations targeting these partitions are routed to this proxy via two-sided RPCs (implemented with \texttt{RDMA\_SEND\&RECV}).
This scheme avoids a large portion of CN-to-MN \texttt{RDMA\_CAS} operations, alleviating network congestion at MN RNICs. 

\vspace{-0.1cm}
\subsection{One-Sided Read vs. Local Read}

Figure~\ref{fig-pre-cas-tpt} and~\ref{fig-pre-read-tpt} also show the performance of \texttt{RDMA\_READ} and \texttt{LOCAL\_READ} (each triggering a single \texttt{memcpy()}), all at 128B granularity. 
Notably, \texttt{LOCAL\_READ} achieves $38.2\times$ higher throughput than \texttt{RDMA\_READ}, 
suggesting that caching KV pairs locally could allow read requests to complete with much higher efficiency. However, as discussed in \S\ref{sec-bg-cache-deploy}, existing memory-disaggregated KV stores either do not cache KV pairs or rely on ownership partitioning, both of which limit performance. 
To address this issue, proxies in \textsf{FlexKV}, in addition to handling RPCs, also maintain directory-based caches for KV pairs. This allows any CN to cache any KV pair, facilitating clients to directly serve read requests from the local cache without triggering any remote operations.

\section{The \textsf{FlexKV} Design}

\subsection{Overview}\label{section_design_overview}

\begin{figure}[tp]
    \centering
    \setlength{\abovecaptionskip}{0.1em}
    \includegraphics[width=\linewidth]{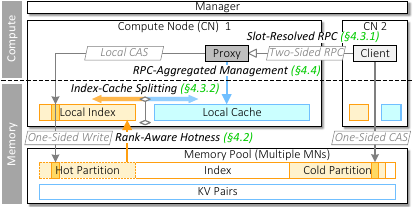}
    \captionsetup{font=small}
    \caption{The overview of \textsf{FlexKV}.}
    \label{fig-design-overview}
\end{figure}

We present \textsf{FlexKV}, a memory-disaggregated KV store with index proxying.
The main idea of index proxying is to dynamically offload the index to CNs, with each CN acting as a proxy responsible for managing an exclusive portion of the global index, leveraging powerful CN CPUs to improve index processing performance and enable efficient compute-side caching.
Figure~\ref{fig-design-overview} illustrates the overview of \textsf{FlexKV}. In each CN, \textit{clients} represent upper-layer applications accessing the KV store, the \textit{proxy} handles RPCs for the \textit{local index}, which is loaded from MNs and consists of several index partitions, and the \textit{local cache} caches the global index and KV pairs. The \textit{memory pool} consists of multiple MNs, holding the index and KV pairs.
The \textit{manager} coordinates cluster-level operations such as hotness detection and parameter tuning.

\noindent \textbf{Request Processing.} 
\textsf{FlexKV} provides a standard KV store API~\cite{fusee, clover, memcached, ramcloud, swarm, aceso}, including \texttt{INSERT}, \texttt{UPDATE}, \texttt{SEARCH}, and \texttt{DELETE}.
We refer to \texttt{SEARCH} as read requests and \texttt{INSERT}, \texttt{UPDATE}, \texttt{DELETE} as write requests. 
The processing of a KV request is broadly divided into two parts: KV pair access and index processing. 
For KV pair access, clients directly read from or write to MNs via one-sided operations. 
For index processing, \textsf{FlexKV} employs a hybrid approach (Figure~\ref{fig-design-overview}): if the corresponding index partition resides on MNs, the index processing is performed directly by the client using one-sided operations, following existing work~\cite{fusee, aceso}. 
Otherwise, if the partition is offloaded to a CN (\textit{i.e.}, index proxying), the index processing is forwarded to the CN's proxy via two-sided RPCs. 
The proxy first writes the index updates to MNs via \texttt{RDMA\_WRITE} to ensure recoverability, then commits the updates to its local index via \texttt{LOCAL\_CAS}. 
The full protocol and correctness analysis of \textsf{FlexKV} is presented in \S\ref{sec-design-together-op}.

\noindent \textbf{RPC Types.} In \textsf{FlexKV}, proxies handle two-sided RPCs for different purposes. RPCs sent from clients to proxies for executing index operations are referred to as index RPCs, which are the primary RPCs in our design. Other types of RPCs are used for auxiliary tasks (\textit{e.g.}, parameter tuning).

\noindent \textbf{Challenges.} 
As analyzed in \S\ref{section_1_intro}, applying index proxying confronts several challenges. To address load imbalance, \textsf{FlexKV} adopts a \textit{rank-aware hotness detection} algorithm to evenly assign index partitions to CNs (\S\ref{section_design_proxy}). To overcome limited CN memory, \textsf{FlexKV} employs a \textit{two-level CN memory optimization} scheme: it uses \textit{slot-resolved index RPCs} to reduce the space overhead of the local index (\S\ref{section_design_index_resolve}), and uses an \textit{adaptive index-cache splitting} technique to partition CN memory between the local index and local cache (\S\ref{section_design_chain}). To reduce cache coherence overhead, \textsf{FlexKV} leverages an \textit{RPC-aggregated cache management} mechanism (\S\ref{section_design_caching}).

\subsection{Rank-Aware Hotness Detection}\label{section_design_proxy}
To enable index proxying, the first step is to assign index partitions to CNs. However, this poses a challenge: many workloads exhibit significant skewness~\cite{pegasus, netcache, facebookworkload, twitter-workload, ycsbworkload}, causing index partitions to experience vastly different access frequencies. With static or random assignment, hot partitions can overload certain CNs while leaving others underutilized, degrading overall performance. Moreover, workload patterns are dynamic in production environments, and frequent reassignments incur non-trivial overhead, making this problem particularly challenging (\textit{\textbf{Challenge 1}}).

To address this challenge, \textsf{FlexKV} adopts a \textit{rank-aware hotness detection} algorithm, as shown in Algorithm~\ref{algorithm-partition}. 
The design is guided by two key strategies. 
First, to ensure load balancing across CNs, it employs a rank-based partition assignment scheme. Second, to prevent oscillations caused by minor hotness fluctuations, it calculates a rank-level displacement score.

\begin{figure}[tp]
    \centering
    \setlength{\abovecaptionskip}{0.1em}
    \includegraphics[width=\linewidth]{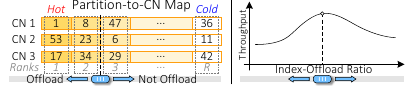}
    \captionsetup{font=small}
    \caption{The rank-based index partition assignment. }
    \label{fig-design-knob}
\end{figure}

\noindent \textbf{Rank-Based Partition Assignment.} 
\textsf{FlexKV} separates the global index into $P = 2^x$ partitions using the first $x$ bits of the key hash (\textit{i.e.}, $x = 13$ in our implementation) and tracks the hotness of each partition. 
To distribute these partitions evenly, the manager sorts all partitions in descending order of hotness (Line~\ref{code_sort_partition}). Let $C$ be the number of CNs. We group these sorted partitions into $R = P/C$ contiguous ranks, each containing $C$ partitions. Rank 1 contains the $C$ hottest partitions, rank 2 contains the next $C$ hottest, and so on. Each CN is then assigned exactly one partition from each rank. Consequently, this assignment forms a partition-to-CN map (Figure~\ref{fig-design-knob}, left), where each CN maintains a hot-to-cold partition list ordered by rank. When a proxy selects partitions to offload based on the memory budget, it scans this list from head to tail, so the hottest partitions are offloaded first. 
This assignment not only balances the global load but also facilitates the \textit{adaptive index-cache splitting} technique in \S\ref{section_design_chain}, where proxies dynamically adjust the number of partitions to offload.

\noindent \textbf{Rank-Level Displacement Score.} 
\textsf{FlexKV} is designed to react only to significant workload shifts. We define the displacement score $D$ to quantify the degree of change in the global hotness distribution. Let $R_{new}(p) \in \{1,\ \ldots,\ R\}$ denote the rank of partition $p$ under the currently observed hotness ordering (Line~\ref{code_new_rank}), and let $R_{old}(p)$ denote its rank under the previous ordering. The rank-level displacement score is then calculated as the sum of the rank changes for all partitions: $D = \sum_{p=1}^P |R_{new}(p) - R_{old}(p)|$.
This metric is appropriate because same-rank partitions share the same offloading priority level. Thus, minor hotness fluctuations that do not change a partition's rank can be safely ignored.

Combining the two strategies above, the manager executes Algorithm~\ref{algorithm-partition} every $\Delta$ seconds (\textit{i.e.}, $\Delta=1$ in our implementation). 
To track hotness, clients maintain a 4-byte per-partition access counter over a $\Delta$-second sliding window. Every $\Delta$ seconds, the manager \texttt{RDMA\_READ}s these counters from CNs and aggregates them to obtain the global hotness (Line~\ref{code_get_hotness}).
To determine when to trigger a reassignment, we compare $D$ against a baseline $B$, which represents the expected displacement of a random reshuffle. Let $X$ and $Y$ be independently uniform on $\{1,\ \ldots,\ R\}$. The per-partition expected rank displacement under a random reshuffle is $E[\,|X-Y|\,] = (R^2 - 1) / (3R)$, a classical formula in mathematics. Hence, the baseline for all partitions is $B = P \cdot E[\,|X - Y|\,] = C (R^2 - 1) / 3$. We trigger reassignment when the current ranking shifts by at least $25\%$ of a fully random reshuffle (Line~\ref{code_trigger_reassign}), \textit{i.e.}, $D \ge 0.25 B$. 
The 25\% threshold is determined empirically to offer a good trade-off between stability and responsiveness. 

\noindent \textbf{Atomic Partition Reassignment.}\label{section_design_atomic_reassign}
To prevent any partition from being served by multiple nodes simultaneously during reassignment, \textsf{FlexKV} uses a two-phase pause-resume protocol. Each CN keeps two partition-to-CN maps: an active map and a staging map that holds the new assignment.
\textit{(1) Pause.} The manager installs the new assignment into all staging maps via \texttt{RDMA\_WRITE}, then notifies all CNs through two-sided RPCs to initiate pausing. Upon receiving the RPC, each CN compares its active and staging maps to identify the partitions whose ownership changes, pauses serving new requests for these partitions, and clears the local cache to ensure coherence. After quiescing these partitions, it returns an ACK.
\textit{(2) Resume.} Once all ACKs arrive, the manager sends resume notifications. Each CN switches its staging map to active, loads newly assigned partitions from MNs, and resumes service for the affected partitions.
In our experiment (\S\ref{section_exp_dynamic_workload}), each reassignment round finishes within 3-5 ms, well below typical multi-second workload shift timescales~\cite{pegasus, netcache, facebookworkload, twitter-workload}.

\begin{algorithm}[tbp]
\small
\caption{Rank-aware hotness detection.}\label{algorithm-partition}
\begin{algorithmic}[1]
\State \textcolor[RGB]{105, 105, 105}{\textit{// $P$: number of partitions}}
\State \textcolor[RGB]{105, 105, 105}{\textit{// $C$: number of CNs}}
\State \textcolor[RGB]{105, 105, 105}{\textit{// access\_count[p][c]: partition p's access count at CN c}}

\State \textcolor[RGB]{105, 105, 105}{\textit{// $R_{\text{old}}[p]$: previous rank of partition $p$}}

\Procedure{\textcolor{blue}{HotnessDetect}}{$P,\ C,\ access\_count,\ R_{old}$} 
    \For{$p \in \{1..P\}$} \textcolor[RGB]{0, 175, 0}{\Comment{\textit{Collect access counts}}}
        \State $hotness[p] \gets \sum_{c=1}^{C} access\_count[p][c]$ \label{code_get_hotness}
    \EndFor

    \State $sorted\_partition \gets$ \textcolor{blue}{\textsc{SortPartition}}($hotness$) \label{code_sort_partition}

    \State $R \gets P / C$ 
    \For{$r \in \{1..R\}$} \textcolor[RGB]{0, 175, 0}{\Comment{\textit{Assign new ranks}}}
        \For{$i \in \{(r-1)C+1 .. rC\}$}
            \State $p \gets sorted\_partition[i]$
            \State $R_{\text{new}}[p] \gets r$ \label{code_new_rank}
        \EndFor
    \EndFor

    \State $D \gets \sum_{p=1}^{P} |R_{\text{new}}[p] - R_{\text{old}}[p]|$ \textcolor[RGB]{0, 175, 0}{\Comment{\textit{Compute displacement}}}

    \State $B \gets C (R^2 - 1)/3$ \textcolor[RGB]{0, 175, 0}{\Comment{\textit{Compute baseline}}}

    \If{$D \ge 0.25B$} \textcolor[RGB]{0, 175, 0}{\Comment{\textit{Trigger reassignment if needed}}} \label{code_trigger_reassign}
        \State \textcolor{blue}{\textsc{TriggerReassignment}}($\,$)
        \State $R_{\text{old}} \gets R_{\text{new}}$
    \EndIf
\EndProcedure
\end{algorithmic}
\end{algorithm}

\subsection{Two-Level CN Memory Optimization}
In the DM architecture, CNs have abundant CPUs but very limited memory. A CN typically has only about 10\% of the memory of a single MN~\cite{clover, dilos, dex}, or even less~\cite{dinomo, chimeluo, sherman}. Index proxying offloads part of the index from MNs to CNs increases pressure on the already limited CN memory and can even degrade performance (\textit{\textbf{Challenge 2}}).

To address this challenge, \textsf{FlexKV} employs the \textit{two-level CN memory optimization} scheme. At the first level, \textit{slot-resolved index RPCs} are used to reduce the inherent space overhead of the local index. At the second level, an \textit{adaptive index-cache splitting} technique is applied to efficiently partition CN memory between the local index and local cache.

\subsubsection{Slot-Resolved Index RPC}\label{section_design_index_resolve}

In many DM index designs, the index alone cannot pinpoint a key's exact index slot without reading the corresponding KV pair. For example, SMART~\cite{smartluo} and RACE~\cite{racehash} store only partial keys or fingerprints, so a key identifies only candidate index slots. The system must fetch the KV pairs referenced by those slots from MNs to confirm the correct one before performing updates. Consequently, in \textsf{FlexKV}, a proxy cannot resolve a key's index slot from the key when serving index RPCs. An alternative is to maintain a direct mapping from keys to slot addresses within the CN's local index, but this map requires storing keys. With variable-sized keys (\textit{e.g.}, strings), the memory overhead becomes substantial, making the local index prohibitively memory-intensive.

To make the local index lightweight, \textsf{FlexKV} issues slot-resolved RPCs. This design is enabled by the fact that, for fault tolerance, the proxy synchronizes every local index update to the MNs' index (\S\ref{sec-design-together-op}). This allows clients to resolve a key's index slot address at the MNs before sending index RPCs to the proxy, thereby avoiding the proxy's own slot lookup and the memory-intensive key-to-slot map. Furthermore, to reduce index lookup traffic to MNs, \textsf{FlexKV} employs different index-resolution strategies for write and read requests.

\begin{figure}[tp]
    \centering
    \setlength{\abovecaptionskip}{0.1em}
    \includegraphics[width=\linewidth]{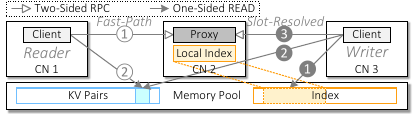}
    \captionsetup{font=small}
    \caption{The slot-resolved index RPC mechanism.}
    \label{fig-design-address-resolve}
\end{figure}

\noindent \textbf{Slot-Resolved Write Requests.} For write requests (\textit{i.e.}, \texttt{INSERT}/\texttt{UPDATE}/\texttt{DELETE}), \textsf{FlexKV} has clients resolve the target index slot before issuing index RPCs to the proxy, as shown in Figure~\ref{fig-design-address-resolve}. Specifically, the client \texttt{RDMA\_READ}s the MNs' index by key to find candidate index slots (\ding{182}), \texttt{RDMA\_READ}s the referenced KV pairs to confirm the correct one (\ding{183}), and then sends that slot address along with the RPC (\ding{184}). The proxy can then update the designated slot directly without a key-based lookup.
Notably, this approach does not add client-side burden and maintains correctness because it follows the original one-sided index protocol exactly, except for replacing the client's \texttt{RDMA\_CAS} with a \texttt{LOCAL\_CAS} at the proxy, while other procedures remain identical.
To accelerate slot resolution, \textsf{FlexKV} embeds the slot address in each key's cache entry (\textit{i.e.}, the KV pair or its address), so the full resolution path is taken only when the local cache misses.

\noindent \textbf{Fast-Path Read Requests.} For read requests (\textit{i.e.}, \texttt{SEARCH}), clients bypass slot resolution and issue index RPCs directly, with the proxy returning all candidate index slots based on the key (\ding{172}). The client then fetches the KV pairs from these slots and finds the one whose key matches (\ding{173}). Since slot collisions are rare, the proxy typically returns only one candidate slot. With this design, most read requests are efficiently served by the proxy, reducing index lookup traffic to MNs.

\subsubsection{Adaptive Index-Cache Splitting}\label{section_design_chain}

Once the local index is introduced on CNs, CN memory must be shared between the local index and the local cache. Under tight CN memory budgets, allocating memory between these two components is challenging. Overprovisioning one component can starve the other, sometimes leading to performance even worse than a simpler design that only has the local cache. Moreover, the optimal memory split is workload-dependent and time-varying (\textit{e.g.}, skew shifts, read-write ratio changes), making static partitioning inherently suboptimal.

To address this issue, \textsf{FlexKV} employs an \textit{adaptive index-cache splitting} technique to search for a near-optimal memory split under dynamic workloads.  
Specifically, it 1) introduces a unified index‑offload ratio across all CNs to reduce the search space, and 2) applies a throughput-guided knob that dynamically tunes this ratio in response to workload shifts.

\noindent \textbf{Unified Index-Offload Ratio.} 
\textsf{FlexKV} uses a sampling-based approach that adjusts the memory split based on observed throughput, offering robustness and implementation simplicity. However, a naive sampler can be slow to converge due to the large search space: with $C$ CNs, there are $C$ split variables to tune. 
To collapse the search space to a single variable, \textsf{FlexKV} introduces a unified index-offload ratio, defined as the ratio between offloaded partitions and total partitions in the per-CN hot-to-cold partition list (Figure~\ref{fig-design-knob}). All CNs use the same ratio to determine the memory allocated to their local index, with the remaining memory assigned to the local cache.
Since index partitions at the same rank across CNs exhibit similar access frequencies (\S\ref{section_design_proxy}), applying this cluster-wide index-offload ratio naturally balances the load.

\begin{algorithm}[tbp]
\small
\caption{Throughput-guided knob.}\label{algorithm_knob}
\begin{algorithmic}[1]

\Procedure{\textcolor{blue}{Knob}}{$\,$}
    \State $i \gets 0$, $s \gets 1$ \textcolor[RGB]{0, 175, 0}{\Comment{\textit{Initialize knob state}}} \label{code_knob_init}

    \While{true}
        \State \textcolor{blue}{\textsc{StartRound}}($\,$)
        \State wait until \textcolor{blue}{\textsc{WorkloadShift}}($\,$) \label{code_workload_shift}
    \EndWhile
\EndProcedure

\Procedure{\textcolor{blue}{StartRound}}{$\,$}
    \State $i_{best} \gets i$, $T_{best} \gets\ $\textcolor{blue}{\textsc{Sample}}($i, \Delta$), $U_{best} \gets 0$
    \State $T_{first} \gets\ $\textcolor{blue}{\textsc{Sample}}($i + s\cdot \delta, \Delta$)
    \If{$T_{first} < T_{best}$}
        \State $s \gets -s$  \textcolor[RGB]{0, 175, 0}{\Comment{\textit{Flip search direction}}} \label{code_change_direct}
    \EndIf
    \While{$U_{best} < 2$}
        \State $i \gets $ $i + s\cdot \delta$, $T \gets\ $\textcolor{blue}{\textsc{Sample}}($i, \Delta$)
        
        \If{$T \le T_{best}$}
            \State $U_{best} \gets U_{best}+1$
        \Else \textcolor[RGB]{0, 175, 0}{\Comment{\textit{Find new candidate}}}
            \State $i_{best} \gets i$, $T_{best} \gets T$, $U_{best} \gets 0$
        \EndIf
        
    \EndWhile
    \State $i \gets i_{best}$
\EndProcedure
\end{algorithmic}
\end{algorithm}

\noindent \textbf{Throughput-Guided Knob.}
\textsf{FlexKV} tunes the index-offload ratio via a periodic, hill-climbing knob guided by sampled throughput (Algorithm~\ref{algorithm_knob}), leveraging the generally unimodal relationship between index-offload ratio and throughput, as we demonstrate in 
\S\ref{section_exp_impact_ratio}.
Let $i \in [0,1]$ denote the fraction of the index offloaded to CNs. Every $\Delta$ seconds, the manager perturbs $i$ by a fixed step size $\delta$, updating $i = i + s \cdot \delta$ according to the current search direction $s \in \{-1, 1\}$. This knob is stateful: it initializes with $i = 0$ and $s = 1$ (Line~\ref{code_knob_init}), and flips direction at the first probe of each round if throughput immediately degrades (Line~\ref{code_change_direct}). It continues stepping while throughput improves. After reaching a candidate maximum, the round terminates once the next two probes both underperform the current best. 
New rounds are triggered by workload shifts (Line~\ref{code_workload_shift}), \textit{i.e.}, either a change of at least $10\%$ in the read-write ratios (\textit{e.g.}, from 1:9 to 2:8) or a partition reassignment (\S\ref{section_design_proxy}).
In our implementation, we set $\Delta = 1$ and $\delta = 0.1$ (\textit{i.e.}, the ratio is adjusted by $10\%$ per second) to prevent overreacting to transient fluctuations. 
While \textsf{FlexKV} currently optimizes for throughput only, the knob can be extended to incorporate other SLOs, such as latency, which is left for future work.

\subsection{RPC-Aggregated Cache Management}\label{section_design_caching}

\begin{figure}[tp]
    \centering
    \setlength{\abovecaptionskip}{0.1em}
    \includegraphics[width=\linewidth]{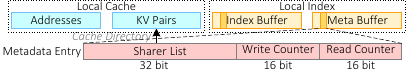}
    \captionsetup{font=small}
    \caption{The memory layout of the compute node.}
    \label{fig-design-caching}
\end{figure}

To accelerate data access, \textsf{FlexKV} implements compute-side caching for both the index and KV pairs, with the latter managed by a directory-based coherence protocol (\S\ref{sec-bg-cache-deploy}). However, this protocol suffers from prohibitive coherence overheads, stemming from both routine metadata maintenance and bursty write invalidations (\textit{\textbf{Challenge 3}}). 

To address this challenge, \textsf{FlexKV} employs an \textit{RPC-aggregated cache management} mechanism. 
The core insight is to leverage the flexibility of two-sided RPCs to manage cache directories and prioritize caching read-intensive KV pairs. 
Unlike one-sided operations that only support simple semantics (\textit{e.g.}, read/write), RPCs handled by CN CPUs can execute arbitrary logic. \textsf{FlexKV} exploits this by piggybacking multiple fine-grained, coherence-related operations (\textit{i.e.}, cache directory updates, KV hotness tracking) onto index RPCs. Since these operations are executed in the same network round-trip as index processing, they incur no additional network overhead and thus reduce the routine metadata maintenance cost. 
Furthermore, the piggybacked KV hotness tracking enables selective caching of read-intensive KV pairs, avoiding caching write-intensive ones that would be quickly invalidated, thus mitigating the impact of bursty write invalidations.

Figure~\ref{fig-design-caching} illustrates the CN memory layout.
The \textit{local cache} is used by clients to accelerate KV request execution, while the \textit{local index} is managed by the proxy. 
For the local cache, it stores either the address or the KV pair of a key, but never both. Cached addresses and KV pairs are managed under a unified FIFO eviction policy, which incurs minimal CPU overhead while providing sufficient performance. 
More advanced eviction strategies are beyond the scope of this paper.
For the local index, it is further divided into the \textit{index buffer} and the \textit{metadata buffer}. The index buffer stores index partitions loaded from MNs, while the metadata buffer holds a metadata entry for each key within these partitions. 
In each metadata entry, the 32-bit \textit{sharer list} bitmap serves as the cache directory that tracks which CNs have cached this KV pair.
For larger-scale clusters, the fixed per-key bitmaps can be replaced with more compact formats (\textit{e.g.}, sparse directory format~\cite{cachespec}), which is orthogonal to our work.
The 16-bit \textit{write counter} and \textit{read counter} increment by 1 when the proxy handles an index RPC from a write request and from a read request, respectively. 
Since 16 bits can easily max out at 65535, the proxy shifts both the write and read counters right by 2 bits when this overflow happens.
Although this reduces precision, it preserves the recent ratio between the write and read counters, which serves as the key indicator for identifying read-intensive KV pairs that merit caching.

\begin{figure}[tp]
    \centering
    \setlength{\abovecaptionskip}{0.1em}
    \includegraphics[width=\linewidth]{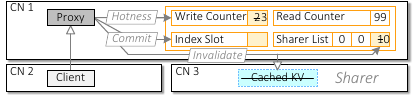}
    \captionsetup{font=small}
    \caption{The RPC-aggregated cache management mechanism. When handling an index RPC for a write request, the proxy increments the write counter, invalidates cached copies via a directory (\textit{i.e.}, the sharer list), and then commits the index slot.}
    \label{fig-design-invalidate}
\end{figure}

\noindent \textbf{Piggybacked Metadata Maintenance.} \textsf{FlexKV} piggybacks numerous coherence-related operations onto index RPCs to reduce the routine metadata maintenance cost, as shown in Figure~\ref{fig-design-invalidate}.
Specifically, when the proxy handles an index RPC, it simultaneously updates the read/write counter of the key to track its read/write hotness, and updates the cache directory to record the sharer\footnote{The \textit{\textbf{sharer}} refers to a CN that holds a cached copy of the KV pair.} if necessary. All of these tasks can be easily finished within the index RPC processing step. Furthermore, since each partition is offloaded to a single CN, all index RPCs for a key are handled by the same CN, meaning the read-write statistics and the cache directory are naturally centralized, eliminating the need for additional aggregation. 

\noindent \textbf{Read-Write Hotness-Aware Caching.} \textsf{FlexKV} selectively caches read-intensive KV pairs to mitigate the impact of bursty write invalidations. When handling an index RPC, in addition to updating the read/write counter, the proxy calculates the ratio between the write and read counters of the key. If this ratio falls below a threshold (\textit{i.e.}, 0.25 in our implementation), the KV pair is considered cache-worthy. The proxy then marks the RPC sender as a sharer in the sharer list, enabling it to cache the KV pair, and notifies the sender via the RPC response to store the KV pair in its local cache.
During the above process, for \texttt{UPDATE} or \texttt{DELETE} requests, the proxy invalidates previous cached copies by issuing two-sided RPCs to CNs in the old sharer list (if non-empty). 
Proxies of these sharer CNs delete their cached copies and check for in-flight RPCs on the same key to prevent them from caching stale KV pairs upon receiving the RPC responses, ensuring consistency in the presence of race conditions.

When a client accelerates a read request through the local cache, it bypasses the proxy (Figure~\ref{fig-design-protocol}), leading to the loss of some read hotness. To address this, clients locally accumulate the read-counter increments when processing read requests without sending index RPCs. The next time an RPC for the same key is issued, the accumulated increments are piggybacked to the proxy for updating. If no RPC is sent for an extended period, or if the accumulated increments exceed a threshold (\textit{i.e.}, 32), the client triggers a dedicated RPC to flush the read-counter increments to the proxy.

\subsection{Put It All Together}\label{section_put_it_all}

\begin{figure}[tp]
    \centering
    \setlength{\abovecaptionskip}{0.1em}
    \includegraphics[width=\linewidth]{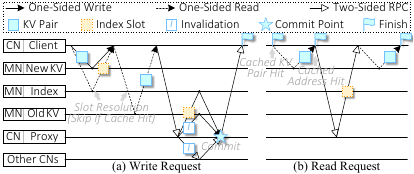}
    \captionsetup{font=small}
    \caption{The workflows of different KV requests in \textsf{FlexKV}.}
    \label{fig-design-protocol}
\end{figure}

This section presents the complete framework of \textsf{FlexKV}, integrating the three previously presented design components.

\noindent \textbf{Index Structure.} 
Without loss of generality, following existing studies~\cite{fusee, aceso, clover, motor, ford}, we use a hash table~\cite{racehash} as a case study for indexing KV pairs.
The hashtable is divided into subtables, which serve as the granularity for index partitioning among proxies in our implementation. Each subtable contains contiguous buckets, each consisting of contiguous 8-byte slots. A slot comprises a 48-bit address field for the KV pair's location in the memory pool, an 8-bit length field for the KV pair size, and an 8-bit fingerprint field for preliminary key matching during lookups. The index uses 8-byte \texttt{CAS} operations to atomically modify index slots.

\noindent \textbf{Memory Management.} To ensure high read performance~\cite{fusee,racehash,clover,aceso}, the out-of-place update scheme is employed for KV pairs, where each write request allocates a new KV pair before modifying the index. To manage KV pair space, \textsf{FlexKV} employs a two-level method~\cite{fusee, aceso}. Clients first request coarse-grained memory blocks (\textit{e.g.}, 16 MB) from MNs when their free space is insufficient, then allocate fine-grained KV pairs from these blocks upon receiving write requests.

\noindent \textbf{Workflows.}\label{sec-design-together-op}
Figure~\ref{fig-design-protocol} presents the \textsf{FlexKV} workflows through which clients execute the four types of KV requests: \texttt{INSERT}, \texttt{UPDATE}, \texttt{SEARCH}, and \texttt{DELETE}. 
As \textsf{FlexKV}'s design mainly targets proxy-managed index partitions, we concentrate on these specific workflows. The workflows for other index partitions follow identical protocols to existing one-sided RDMA based KV stores~\cite{aceso, fusee}, and are omitted for brevity.

\textit{(1) Write.} 
For a write request (\texttt{INSERT}/\texttt{UPDATE}/\texttt{DELETE}), the client first writes the new KV pair to MNs while resolving the index slot address. The resolution can be skipped when the local cache is hit (\S\ref{section_design_index_resolve}). After this, the client issues an index RPC to the proxy, which then invalidates previous caches: (i) for cached addresses, the proxy clears the previously committed KV pair's valid bit to invalidate (\S\ref{sec-bg-cache-deploy}); (ii) for cached KV pairs, the proxy sends RPCs to those CNs on the sharer list to invalidate (\S\ref{section_design_caching}). In parallel, the proxy writes the new index slot to the MNs' index for fault tolerance. Finally, it commits the new slot to its local index via \texttt{LOCAL\_CAS} (\textbf{commit point}) and replies to the client.

\textit{(2) Read.}
For a read request (\texttt{SEARCH}), the client first checks the local cache. If a cached KV pair exists, the client immediately returns the result. If a cached address exists, the client performs an \texttt{RDMA\_READ} to fetch the KV pair from MNs and verifies the match. If both miss, the client sends an index RPC to the proxy to obtain the address of the latest committed KV pair (\S\ref{section_design_index_resolve}), which is then retrieved via \texttt{RDMA\_READ}.

When handling index RPCs for write requests, the proxy uses a local key-to-lock map to ensure only one write request per key is processed at a time. Concurrent writes to keys already in process will fail immediately, as in \texttt{CAS}. Since in-flight RPCs are few, the map size remains negligible.

\noindent \textbf{Fault Tolerance.} 
We assume that the network is reliable and no Byzantine failures~\cite{byzantine} occur, following previous work~\cite{fusee,swarm,aceso,motor}.
\textit{(1) For the manager,} it only performs periodic parameter tuning tasks without maintaining any persistent state. Its failure is safe and does not affect clients.
\textit{(2) For MNs,} currently, \textsf{FlexKV} employs primary-backup replication~\cite{primary-backup} to ensure MN data safety, where each write is replicated to multiple MNs (3-way in our evaluation).
\textit{(3) For CNs,} \textsf{FlexKV} treats all CN data as disposable. Loss of local caches is safe, as they contain temporary data. Loss of local indexes is also safe since proxies write all updates to MNs before modifying the local indexes. 
As MN failures can be effectively handled by existing solutions~\cite{fusee, swarm, hermes-replica, splitft}, we focus on CN failures.

Upon detecting a CN failure (via a membership service~\cite{ukharonmember, zookeeper}), surviving CNs clear their local caches, since the failed CN's cache directory is lost and cached KV pairs it managed can no longer be validated for coherence. These CNs also update their partition-to-CN maps (\S\ref{section_design_proxy}) to redirect all requests targeting the failed CN's index partitions to MNs, allowing clients to continue accessing those partitions via one-sided operations. The entire response completes within 3-5ms in our experiments, ensuring minimal service disruption.
When the failed CN restarts, the manager re-offloads the corresponding index partitions to it using the same procedure as the atomic partition reassignment (\S\ref{section_design_atomic_reassign}).

\noindent \textbf{Linearizability and Correctness.} 
\textsf{FlexKV} guarantees linearizability~\cite{linearizability} by 
treating the proxy's \texttt{LOCAL\_CAS} as the atomic commit point.
We demonstrate correctness by analyzing the visibility of KV pair versions across three possible read paths: \ding{182} cached KV pair, \ding{183} cached address, and \ding{184} index RPC.

\textit{(1) Path convergence.} Before \texttt{LOCAL\_CAS}, the proxy strictly enforces cache invalidations, effectively disabling paths \ding{182} and \ding{183}. This forces all concurrent reads to converge to path \ding{184}, making the proxy the single source of truth.

\textit{(2) Atomic visibility switch.} With read traffic converged, \texttt{LOCAL\_CAS} acts as an atomic boundary. Before execution, the proxy's local index references the old KV pair; immediately after, the new one. Note that the new index slot written to MNs before \texttt{LOCAL\_CAS} is used solely for slot resolution (\S\ref{section_design_index_resolve}) and does not expose the uncommitted write. 

Finally, if a proxy fails after writing the new index slot to MNs but before executing \texttt{LOCAL\_CAS}, clients fall back to accessing MNs directly, making the uncommitted write visible. However, this can be treated as a redo of an incomplete write~\cite{fusee}, preserving linearizability.

\noindent \textbf{Garbage Collection.}
\textit{(1) For KV pairs.} Since \textsf{FlexKV} employs
out-of-place KV pair updates, it needs to reclaim stale KV pairs. When clients execute \texttt{UPDATE} or \texttt{DELETE} requests, they add the old KV pairs to a free list in each CN, allowing their space to be reused by subsequent allocations, similar to previous work~\cite{clover}. 
\textit{(2) For index slots.} \texttt{DELETE} frees a key's index slot, but other CNs may still hold cached slot addresses for this key (\S\ref{section_design_index_resolve}). \textsf{FlexKV} uses a lease-based method to ensure safety: each cached slot address is valid for at most $T_{lease}$ time and is renewed on each successful operation on that key. We repurpose the 48-bit address field in index slots: the first bit serves as a valid bit, while the remaining 47 bits store either the actual address (when valid=1) or a \texttt{DELETE} timestamp $T_{delete}$ (when valid=0). During \texttt{DELETE}, instead of clearing the slot, we set the valid bit to 0 and record $T_{delete}$. Subsequent \texttt{INSERT} can safely reuse an invalid slot once the current time exceeds $T_{delete} + T_{lease} \times (1 + \delta)$, where $\delta$ accounts for the bounded cluster-wide clock drift (up to $\delta = 10^{-4}$~\cite{shiftlock, sundial, graham}). In our implementation, we set $T_{lease} = 200$ ms.

\noindent \textbf{Other Optimizations.}
\textsf{FlexKV} incorporates several optimizations: \textit{(1) UD-based RPC.} We use unreliable datagram (UD) QPs for two-sided RPCs to avoid QP explosion between clients and proxies, with timeouts ensuring reliability~\cite{fasst}. \textit{(2) RDMA optimizations.} We adopt standard techniques including coroutines, inline writes, doorbell batching, and selective signaling~\cite{fusee, sherman, designguideatc}. To ensure fairness, we also add these optimizations to all baselines in our evaluation (\S\ref{section_evaluation}).

\subsection{Discussion}

\noindent \textbf{Generality of \textsf{FlexKV} to Other KV Stores.}
The design of \textsf{FlexKV} is general-purpose.
First, offloading index partitions to CNs is applicable to other index data structures, \textit{e.g.}, SMART~\cite{smartluo}, Sherman~\cite{sherman}, and CHIME~\cite{chimeluo}, since one-sided RDMA operations can be directly replaced with local memory accesses.
Once the index is offloaded, the proposed compute-side caching and memory partitioning designs become readily applicable to these systems.

\noindent \textbf{Compatibility of \textsf{FlexKV} over CXL.}\label{section_discuss_cxl} 
We consider CXL 3.0, which enables hardware-based memory sharing similar to RDMA. Although CXL 3.0 provides one-sided atomics via a MESI-like coherence protocol~\cite{cxlmesi,cxlspec}, recent studies reveal performance challenges~\cite{cxldemysify,cxlhydrarpc,cxlshm,cxltrain}. Thus, \textsf{FlexKV}'s techniques remain valuable to future CXL-based DM. For inter-CN RPCs, \textsf{FlexKV} can leverage CXL-based RPC designs, as explored in some pioneering work~\cite{cxlhydrarpc,cxltelerpc}.
\section{Evaluation}\label{section_evaluation}
\subsection{Experimental Setup}\label{sec-exp-setup}

\begin{figure*}[tbp]
\centering
\subfigure[YCSB A]{
    \includegraphics[width=0.24\textwidth]{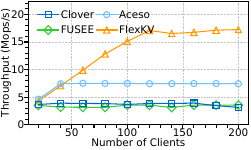}
}
\hspace{-0.8em}
\subfigure[YCSB B]{
    \includegraphics[width=0.24\textwidth]{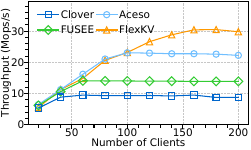}
}
\hspace{-0.8em}
\subfigure[YCSB C]{
    \includegraphics[width=0.24\textwidth]{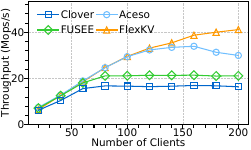}
}
\hspace{-0.8em}
\subfigure[YCSB D]{
    \includegraphics[width=0.24\textwidth]{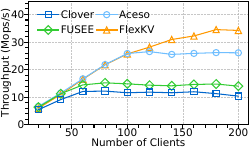}
}
\hspace{-0.8em}
\captionsetup{font=small}
\caption{The throughput under YCSB workloads.}
\label{fig-exp-ycsb-tpt}
\end{figure*}

\begin{figure*}[tbp]
\centering
\subfigure[YCSB A]{
    \includegraphics[width=0.24\textwidth]{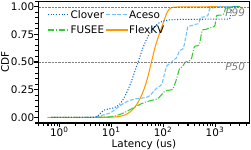}
}
\hspace{-0.8em}
\subfigure[YCSB B]{
    \includegraphics[width=0.24\textwidth]{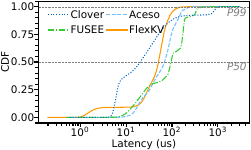}
}
\hspace{-0.8em}
\subfigure[YCSB C]{
    \includegraphics[width=0.24\textwidth]{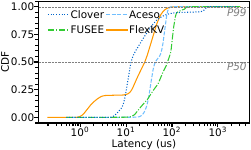}
}
\hspace{-0.8em}
\subfigure[YCSB D]{
    \includegraphics[width=0.24\textwidth]{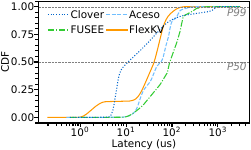}
}
\hspace{-0.8em}
\captionsetup{font=small}
\caption{The latency CDF under YCSB workloads.}
\label{fig-exp-ycsb-lat}
\end{figure*}

\noindent \textbf{Testbed.} We conduct experiments with 23 physical machines (\textit{i.e.}, 20 CNs and 3 MNs) on the Apt cluster of CloudLab~\cite{cloudlab}, with the manager colocated on the first CN. Each machine features a 56Gbps Mellanox ConnectX-3 IB RNIC connected to a 56Gbps Mellanox SX6036G switch, two 8-core Intel E5-2650v2 CPUs, and 64GB of DRAM. We present the sensitivity results for different CN-MN ratios in \S\ref{section_exp_sensitivity}.

\noindent \textbf{Comparisons.} 
We compare \textsf{FlexKV} with three state-of-the-art memory-disaggregated KV stores: Clover~\cite{clover}, FUSEE~\cite{fusee}, and Aceso~\cite{aceso}. Clover places the index on additional monolithic servers (\textit{i.e.}, metadata servers), requiring separate accesses for index operations (Figure~\ref{fg-bg-index}(a)). To ensure fairness, we allocate an additional physical machine to Clover as the metadata server. FUSEE and Aceso, conversely, place the index in MNs, with clients performing index operations via one-sided RDMA (Figure~\ref{fg-bg-index}(b)). The key difference is that FUSEE employs replication for index fault tolerance, requiring multiple \texttt{RDMA\_CAS} to multiple replicas for each index update, while Aceso adopts a checkpointing mechanism that requires only a single \texttt{RDMA\_CAS}. 
To demonstrate the impact of ownership partitioning~\cite{dinomo, dex}, we implement \textsf{FlexKV-OP}, a variant of \textsf{FlexKV} adopting the ownership partitioning scheme. In \textsf{FlexKV-OP}, each request must first be forwarded to a designated CN, determined by hashing, for processing.

\noindent \textbf{Workloads.} 
We employ YCSB benchmarks~\cite{ycsbworkload} and Twitter workloads~\cite{twitter-workload} for our evaluation.
The YCSB benchmarks use four core workloads: A (write-intensive, 50\% \texttt{UPDATE} and 50\% \texttt{SEARCH}), B (read-intensive, 5\% \texttt{UPDATE} and 95\% \texttt{SEARCH}), C (read-only, 100\% \texttt{SEARCH}), and D (insert-involved, 5\% \texttt{INSERT} and 95\% \texttt{SEARCH}). 
By default, all YCSB workloads employ skewed key generation following a Zipfian distribution ($\alpha=0.99$\footnote{The parameter $\alpha$ is also called \textbf{\textit{theta}} in some papers~\cite{zipfian-generate}.})~\cite{zipfian}, as per the YCSB standard configuration. 
We also evaluate the performance under uniform YCSB workloads in \S\ref{section_exp_ycsb_uniform}.
The Twitter workloads comprise 54 real-world traces collected from production clusters, with varying read-write ratios, KV pair sizes, and skewness.

\noindent \textbf{Parameters.} Unless otherwise specified, we use 200 clients evenly distributed across 20 CNs, with each client running 8 coroutines. Each MN provides 56~GB of memory, for a total of 168~GB. All systems use 3-way replication or an equivalent fault-tolerance setup.
Each CN has 64~MB of memory, roughly 5\% of the working-set size, matching DM's configuration. 
For all YCSB experiments, we bulk-load 10 million KV pairs before issuing any requests, and the KV pair size is set to 128~B, which is representative of real-world workloads~\cite{twitter-workload, rocksdb, ycsbworkload, facebookworkload}. 
Sensitivity results for varying KV pair sizes and CN memory limits are in \S\ref{section_exp_sensitivity}.
To ensure fairness, points with the same index across curves in a figure use identical configurations (the number of CNs, clients, and coroutines).

\subsection{Performance Comparison}

\noindent \textbf{YCSB Skewed Workloads.} 
We present the performance of \textsf{FlexKV}, Aceso, FUSEE, and Clover under YCSB workloads. To generate throughput-client curves, we vary the number of clients from 20 to 200.
The results are shown in Figure~\ref{fig-exp-ycsb-tpt}.
\textsf{FlexKV} achieves the highest throughput across all workloads, with peak throughput improvements over the second-best system of $2.31\times$ (YCSB A), $1.34\times$ (YCSB B), $1.37\times$ (YCSB C), and $1.31\times$ (YCSB D).
Aceso and FUSEE are limited by their reliance on expensive \texttt{RDMA\_CAS} for index operations, making MN RNICs a bottleneck. Clover suffers from frequent accesses that overload its metadata server. In contrast, \textsf{FlexKV} leverages index proxying to balance index operations across all nodes and employs KV pair caching to achieve superior performance on different workloads. 

We further evaluate request latency for all systems with 200 clients, presenting the cumulative distribution functions (CDFs) in Figure~\ref{fig-exp-ycsb-lat}. Clover achieves the lowest P50 latency because, while it bottlenecks at the metadata server, clients that hit the address cache bypass the metadata server and access MNs directly. Since Clover's throughput is lower, its MN RNICs experience less congestion and can service RDMA operations quickly. For P99 latency, \textsf{FlexKV} achieves the lowest values across YCSB A/B/C/D, reducing P99 latency by $85.2\%$, $36.4\%$, $4.1\%$, and $36.9\%$ compared to the second-best system, respectively. On YCSB C, \textsf{FlexKV}'s P99 latency is close to Aceso's, as \textsf{FlexKV}'s $37\%$ higher throughput increases network pressure and offsets its latency advantage.

\begin{figure}[tbp]
    \centering
    \begin{minipage}[t]{0.49\linewidth}
      \centering
      \includegraphics[width=\textwidth]{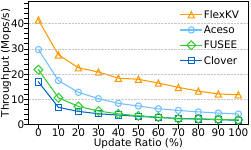}
      \captionsetup{font=small}
      \caption{The throughput under different \texttt{UPDATE}-\texttt{SEARCH} ratios.}
      \label{fig-exp-ycsb-upd-ratio}
    \end{minipage}
    \hspace{0em}
    \begin{minipage}[t]{0.49\linewidth}
      \centering
      \includegraphics[width=\textwidth]{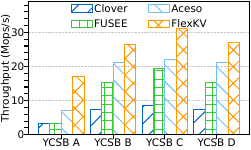}
      \captionsetup{font=small}
      \caption{The throughput under uniform YCSB workloads.}
      \label{fig-exp-uni-ycsb-tpt}
    \end{minipage}
\end{figure}

\noindent \textbf{Different Update-Search Ratios.} 
Figure~\ref{fig-exp-ycsb-upd-ratio} shows the throu-ghput under different \texttt{UPDATE}-\texttt{SEARCH} ratios. Throughput decreases for all systems as the proportion of \texttt{UPDATE} increases, since \texttt{UPDATE} is more complex than \texttt{SEARCH}. \textsf{FlexKV} consistently achieves the highest throughput across all ratios. In read-intensive scenarios, \textsf{FlexKV} mainly benefits from effective KV pair caching, while in write-intensive scenarios, from index proxying that offloads index processing to CNs.

\noindent \textbf{YCSB Uniform Workloads.}\label{section_exp_ycsb_uniform} 
We modify the key distribution in the YCSB workloads from Zipfian to uniform, and Figure~\ref{fig-exp-uni-ycsb-tpt} shows the results. \textsf{FlexKV}'s throughput drops because the uniform access pattern lowers local cache hit ratios. 
Moreover, under the uniform distribution, the offloaded index partitions become less frequently accessed than in the skewed case, which reduces the effectiveness of index proxying.
\begin{figure}[tbp]
    \centering
    \includegraphics[width=\linewidth]{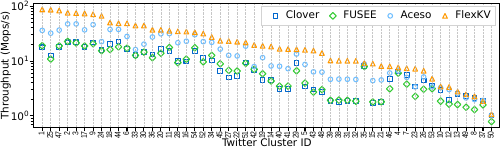}
    \captionsetup{font=small}
    \caption{The throughput under Twitter workloads.}
    \label{fig-exp-twitter-tpt}
\end{figure}

\noindent \textbf{Twitter Workloads.}
Figure~\ref{fig-exp-twitter-tpt} shows the throughput under Twitter workloads, with clusters ordered by \textsf{FlexKV}'s throughput in descending order. 
Across all clusters, \textsf{FlexKV} achieves $1.03$–$2.94\times$ higher throughput than the second-best system. 
For some workloads, the gains are modest, as they are either uniform (\textit{e.g.}, cluster 35 with Zipfian $\alpha=0$) or involve large KV pairs, causing all systems to be limited by network bandwidth (\textit{e.g.}, cluster 50). 
In contrast, workloads with high skewness and read-intensive patterns benefit substantially. For example, cluster 1 ($\alpha=2.68$, 99\% reads) sees significant improvement from \textsf{FlexKV}'s KV pair caching. 
These results highlight \textsf{FlexKV}'s effectiveness in real-world environments.

\subsection{System-level Analysis}

\begin{figure}[tbp]
    \centering
    \begin{minipage}[t]{0.49\linewidth}
      \centering
      \includegraphics[width=\textwidth]{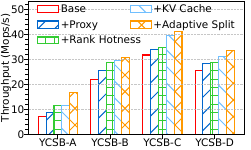}
      \captionsetup{font=small}
      \caption{The ablation study on \textsf{FlexKV}.}
      \label{fig-exp-ablation}
    \end{minipage}
    \hspace{0em}
    \begin{minipage}[t]{0.49\linewidth}
      \centering
      \includegraphics[width=\textwidth]{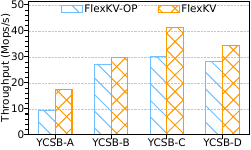}
      \captionsetup{font=small}
      \caption{The effect of ownership partitioning on \textsf{FlexKV}.}
      \label{fig-exp-op}
    \end{minipage}
\end{figure}

\noindent \textbf{Ablation Study.}
Figure~\ref{fig-exp-ablation} shows the impact of \textsf{FlexKV}'s key techniques on performance under YCSB workloads.
\textit{\textbf{Base}} is the basic version with only address caching. \textit{\textbf{+Proxy}} enables the index proxies, statically offloading the first 20\% index partitions to CNs, improving throughput by 14.5\% on average across four workloads. \textit{\textbf{+Rank Hotness}} identifies and offloads the hottest index partitions to CNs, improving throughput by 11.9\% on average. \textit{\textbf{+KV Cache}} enables KV pair caching, improving throughput by 6.1\% on average. \textit{\textbf{+Adaptive Split}} dynamically adjusts the index-offload ratio, improving throughput by 15.2\% on average. Together, these techniques incrementally build up to the full performance of \textsf{FlexKV}.

\noindent \textbf{Cost of Ownership Partitioning.} \label{section_exp_op_cost} 
As mentioned in \S\ref{sec-exp-setup}, we implement \textsf{FlexKV-OP}, a variant of \textsf{FlexKV} that employs ownership partitioning similar to DINOMO~\cite{dinomo} and DEX~\cite{dex}, where each KV request is forwarded to a designated CN based on its key for subsequent processing. This design lets each CN manage a distinct key range, maintaining local caches without coherence overhead. However, as shown in Figure~\ref{fig-exp-op}, ownership partitioning reduces \textsf{FlexKV}'s throughput by 24.9\% on average across four workloads, due to the extra network overhead caused by request forwarding.

\begin{figure}[tbp]
    \centering
    \includegraphics[width=\linewidth]{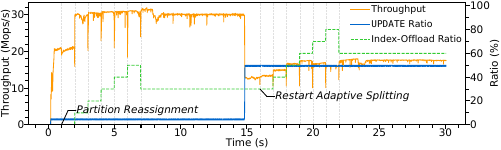}
    \captionsetup{font=small}
    \caption{The throughput under dynamic workload pattern.}
    \label{fig-exp-dynamic}
\end{figure}

\noindent \textbf{Dynamic Workload Pattern.}\label{section_exp_dynamic_workload}
To demonstrate \textsf{FlexKV}'s adaptability to dynamic workload patterns, we run YCSB B (5\% \texttt{UPDATE}) for 15~seconds, then switch to YCSB A (50\% \texttt{UPDATE}), as shown in Figure~\ref{fig-exp-dynamic}. The performance is sampled every 10~ms. At 1~s, the manager runs Algorithm~\ref{algorithm-partition} for index partition reassignment. At 2~s, the manager runs Algorithm~\ref{algorithm_knob} to search for a near-optimal index-offload ratio. At 7~s, it finds the ratio of 30\%. At 8~s, throughput drops slightly as cache evictions increase CPU overhead when local caches fill up. At 15~s, the \texttt{UPDATE} ratio changes to 50\%, causing throughput to drop to 12.5 Mops/s. At 16~s, the manager detects the workload pattern change and re-runs Algorithm~\ref{algorithm_knob} to search for a new index-offload ratio. At 22~s, it finds the new ratio of 60\%, improving throughput to 17.3 Mops/s. This demonstrates \textsf{FlexKV}'s ability to adapt to workload changes and maintain an appropriate index-cache memory split. Throughout the process, each index partition reassignment or index-offload ratio adjustment takes 3.24-5.26~ms, causing minimal disruption.

\begin{figure}[tbp]
    \centering
    \begin{minipage}[t]{0.49\linewidth}
      \centering
      \includegraphics[width=\textwidth]{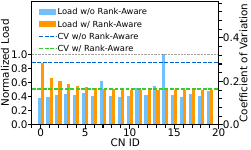}
      \captionsetup{font=small}
      \caption{The load distribution across CNs.}
      \label{fig-exp-load-balance}
    \end{minipage}
    \hspace{0em}
    \begin{minipage}[t]{0.49\linewidth}
      \centering
      \includegraphics[width=\textwidth]{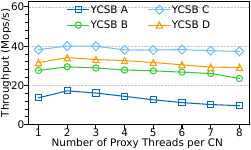}
      \captionsetup{font=small}
      \caption{The throughput under different proxy thread numbers.}
      \label{fig-exp-sense-proxy-num}
    \end{minipage}
\end{figure}

\noindent \textbf{Effect of Load Balancing.} To demonstrate the effect of the \textit{rank-aware hotness‑detection} algorithm on load balancing, Figure~\ref{fig-exp-load-balance} shows the load distribution across CNs under YCSB A and reports the coefficient of variation (CV) as the balancing metric. 
Enabling the Algorithm~\ref{algorithm-partition} increases the total index operations offloaded to CNs by 14.8\% and reduces the CV by 42.7\%, significantly improving load balancing.

\noindent \textbf{Performance Breakdown.} We present the index-offload ratios, cache hit ratios, and average \texttt{SEARCH} latencies with 200 clients in Table~\ref{tab-exp-breakdown}. For write-intensive YCSB A, the index-offload ratio is high (60\%) to reduce the pressure of frequent \texttt{RDMA\_CAS} on MNs, while the value cache hit ratio remains low (0.1\%) as \textsf{FlexKV} avoids caching frequently modified KV pairs. For read-intensive YCSB C, the index-offload ratio reaches 80\% because more index partitions managed by proxies enable more KV pairs to be cached, achieving a higher value cache hit ratio (18.9\%). YCSB B exhibits a lower index-offload ratio (30\%) as it is neither write-intensive nor has good caching locality. The latency breakdown shows that KV-pair cache hits incur only 2$\mu$s due to local data structure overhead (\textit{e.g.}, cache lookup and eviction), while address cache hits require one \texttt{RDMA\_READ} with about 20$\mu$s latency. When both caches miss, latency increases to about 50$\mu$s.

\begin{table}[tbp]
\centering
\footnotesize
\setlength{\tabcolsep}{4pt}
\caption{\small The performance breakdown under YCSB workloads.}
\label{tab-exp-breakdown}
\begin{tabular}{c|c|cc|ccc}
\toprule[0.5pt]
\multirow{2}{*}{\centering Workload\vspace{-5pt}} & \multirow{2}{*}{\centering Offload Ratio (\%)\vspace{-5pt}} &
\multicolumn{2}{c|}{Hit Ratio (\%)} & \multicolumn{3}{c}{\texttt{SEARCH} Latency ($\mu$s)} \\
\cmidrule(lr){3-4} \cmidrule(lr){5-7}
 & & KV & Addr & KV & Addr & Other \\
\midrule[0.5pt]
YCSB A & 60 & 0.1 & 10.4 & 2.3 & 24.1 & 54.1 \\
YCSB B & 30 & 10.1 & 24.1 & 1.9 & 23.6 & 52.3 \\
YCSB C & 80 & 18.9 & 30.6 & 2.2 & 16.5 & 42.8 \\
YCSB D & 50 & 15.5 & 31.3 & 2.3 & 23.3 & 47.4 \\
\bottomrule[0.5pt]
\end{tabular}
\end{table}

\subsection{Sensitivity Analysis}\label{section_exp_sensitivity}

\noindent \textbf{Impact of Proxy Thread Number.} 
In \textsf{FlexKV}, each proxy runs multiple threads to process index RPCs concurrently. Figure~\ref{fig-exp-sense-proxy-num} shows throughput under YCSB workloads with different numbers of proxy threads. Throughput first increases with the number of threads, peaks at 2 threads, and then declines due to resource contention (\textit{e.g.}, lock contention on multithreaded data structures and RNIC cache thrashing from more QPs~\cite{rdmaunderstand, fasst, smartren}). We set the default number of proxy threads to 2. Considering CNs' abundant CPUs and \textsf{FlexKV}'s up to 2.94$\times$ performance improvement, this modest 20\% extra CPU usage is acceptable. Notably, a single proxy thread already achieves 71.8–90.2\% of the peak throughput, indicating the CPU overhead of index proxying is small.

\begin{figure}[tbp]
    \centering
    \begin{minipage}[t]{0.49\linewidth}
      \centering
      \includegraphics[width=\textwidth]{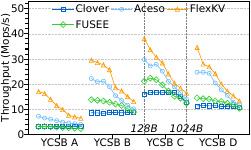}
      \captionsetup{font=small}
      \caption{The throughput under different KV pair sizes.}
      \label{fig-exp-sense-kv-size}
    \end{minipage}
    \hspace{0em}
    \begin{minipage}[t]{0.49\linewidth}
      \centering
      \includegraphics[width=\textwidth]{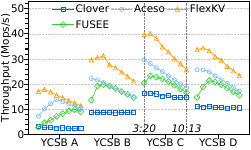}
      \captionsetup{font=small}
      \caption{The throughput under different CN-MN ratios.}
      \label{fig-exp-sense-cn-mn-ratio}
    \end{minipage}
\end{figure}

\noindent \textbf{Impact of KV Pair Size.} 
Figure~\ref{fig-exp-sense-kv-size} shows the throughput under KV pair sizes ranging from 128B to 1024B, increasing in 128B steps. As the KV size grows, throughput of all systems gradually decreases and their performance gap narrows due to bandwidth saturation. Nevertheless, \textsf{FlexKV} consistently achieves the highest throughput among the four systems.

\begin{figure}[tbp]
    \centering
    \begin{minipage}[t]{0.49\linewidth}
      \centering
      \includegraphics[width=\textwidth]{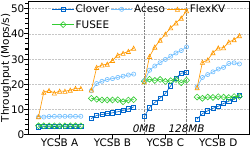}
      \captionsetup{font=small}
      \caption{The throughput under different CN memory limits.}
      \label{fig-exp-sense-cn-mem}
    \end{minipage}
    \hspace{0em}
    \begin{minipage}[t]{0.49\linewidth}
      \centering
      \includegraphics[width=\textwidth]{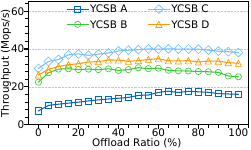}
      \captionsetup{font=small}
      \caption{The throughput under different index-offload ratios.}
      \label{fig-exp-sense-offload-ratio}
    \end{minipage}
\end{figure}

\noindent \textbf{Impact of CN-MN Ratio.} 
While previous experiments use only 3 MNs, here we vary the CN-MN ratio from 3:20 to 10:13 to evaluate \textsf{FlexKV}'s scalability, as shown in Figure~\ref{fig-exp-sense-cn-mn-ratio}.
Throughput of \textsf{FlexKV}, Aceso, and FUSEE first increases and then decreases. The initial increase is due to more MN RNICs, which mitigate the MN-side network bottleneck. The subsequent decrease occurs because each client reaches its CPU limit, and fewer CNs reduce the total number of clients. For Clover, performance remains limited by the metadata server, so varying CN-MN ratios has little effect.

\noindent \textbf{Impact of CN Memory Limit.}  
Figure~\ref{fig-exp-sense-cn-mem} shows the throughput under CN memory limits ranging from 0MB to 128MB, increasing in 16MB steps. 
Throughput generally increases with larger CN memory, especially for read-intensive workloads. For the write-intensive YCSB A, only \textsf{FlexKV} shows significant improvement, as the other systems are limited by \texttt{RDMA\_CAS} or the metadata server. FUSEE shows little improvement because it prefetches index hash buckets even when cached addresses hit to reduce RTT, which causes read-amplification and leaves FUSEE bandwidth-bound. In contrast, Aceso only fetches buckets on cache misses~\cite{aceso}.

\noindent \textbf{Impact of Index-Offload Ratio.}\label{section_exp_impact_ratio} 
We disable \textsf{FlexKV}'s knob (Algorithm~\ref{algorithm_knob}) to fix the index-offload ratio, and manually vary it to evaluate performance. As shown in Figure~\ref{fig-exp-sense-offload-ratio}, under YCSB workloads, throughput first increases and then slightly decreases as the offload ratio grows. This indicates a roughly unimodal relationship between the index-offload ratio and throughput, motivating Algorithm~\ref{algorithm_knob} to find a near-optimal ratio. It also shows that placing the index entirely in MNs or fully offloading it to CNs is suboptimal, while \textsf{FlexKV}'s partial offloading strategy achieves the best performance by fully utilizing all hardware resources in the cluster (Figure~\ref{fg-bg-index}).

\section{Related Work}

\noindent \textbf{Disaggregated Memory.}
Existing studies on DM include software-based approaches, such as modifying operating systems~\cite{legoos, fastswap, hermit, literdma, krcore, canvas, taleoftwo} and runtimes~\cite{semeru, memliner, mira, finemem, rethink} for seamless integration, as well as efforts to optimize data structures~\cite{aifm, carbink, citron, cxlshm, chimeluo, sherman} and specific applications like in-memory KV stores~\cite{dinomo, dex, swarm, aceso, lolkv, fusee} and transactional storage systems~\cite{motor, ford, hdtx, hybrid-better}. On the other hand, hardware-based approaches~\cite{dmorigin, cxldemysify, cxlpond, cxltpp} and hardware-software co-designed methods~\cite{mind, clio, concordia} are being explored to enhance functionality. \textsf{FlexKV}, an advanced KV store designed for DM, can also benefit from these low-level optimizations. 

\noindent \textbf{Memory-Disaggregated KV Stores.}
Recently, many KV stores targeting DM architectures have emerged, such as Clover~\cite{clover}, FUSEE~\cite{fusee}, DINOMO~\cite{dinomo}, Aceso~\cite{aceso}, and SWARM-KV~\cite{swarm}, each with its unique design focus: Clover reduces monetary and energy costs, FUSEE disaggregates metadata management, DINOMO leverages ownership partitioning for cache efficiency, Aceso uses erasure coding to save space, and SWARM-KV trades space for one-RTT operations. 
Finally, \textsf{FlexKV} focuses on scalable index processing and flexible KV pair caching to improve performance.

\noindent \textbf{Memory-Disaggregated Index Processing.}
In memory-disaggregated KV stores, index processing are handled in various ways. Clover~\cite{clover} relies on monolithic metadata servers, which limits resource efficiency and can become the performance bottleneck. Aceso~\cite{aceso} and FUSEE~\cite{fusee} let CN CPUs manage the index in MNs remotely via one-sided RDMA. DEX~\cite{dex} also uses one-sided RDMA but can offload some operations to MN CPUs for local execution, requiring MN compute power. DINOMO~\cite{dinomo} relies entirely on MN CPUs, demanding significant MN compute power. In contrast, \textsf{FlexKV} allows CN CPUs to either manage the index remotely via one-sided RDMA or act as the proxies for local execution, without imposing any compute overhead on MNs.

\noindent \textbf{Memory-Disaggregated Caching.}
Ditto~\cite{ditto} is a memory-disaggregated caching system for backup storage. It stores cached data in MNs and performs distributed eviction via one-sided RDMA. In contrast, \textsf{FlexKV} stores cached data in CNs and performs local eviction via CN CPUs. DiFache~\cite{difache}, a concurrent work with \textsf{FlexKV}, is a general-purpose caching system for DM applications. It stores cached data in CNs but keeps cache directory in MNs, which is managed via one-sided RDMA. In contrast, \textsf{FlexKV} stores both cached data and cache directory in CNs, and directory management is piggybacked on index RPCs, reducing extra network overhead.

\section{Conclusion}

This paper presents \textsf{FlexKV}, a memory-disaggregated KV store with index proxying. \textsf{FlexKV} offloads the index to CNs dynamically and introduces KV pair caching across CNs, 
addressing the bottleneck at MN RNICs. Evaluation shows it improves throughput by up to $2.94\times$ and reduces latency by up to 85.2\%, compared with the state-of-the-art systems. 

\bibliographystyle{plain}
\bibliography{FlexKV}

\end{document}